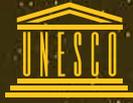
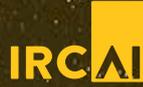

United Nations
Educational, Scientific and
Cultural Organization

International Research Centre
on Artificial Intelligence
under the auspices of UNESCO

*Opinion Series Report*
**An AI-based Learning Companion Promoting Lifelong Learning Opportunities for All**

*Note from the editor*

*Throughout 2020 the IRCAI programme committees authored an Opinion Series on a range of topics across AI and sustainability as part of a wider portfolio of IRCAIs international initiatives. This opinion series explores what AI currently means to researchers across the world and across a variety of disciplines both in AI and sustainable development, and was led by the IRCAI executive. The views and opinions expressed by authors are their own and do not reflect the position of IRCAI, but are simply an illustration of the various opinions reflective of the diverse initiatives that IRCAI is pursuing, from basic research to policy creation in and around AI.*

*As IRCAI is working from home for the foreseeable future, it means we arenot organising in person events. Introducing this new quick report format has the same spirit — original stripped-down research pieces, an attempt to an intellectual dialogue — just a different space.*

*Davor Orlic, Chief Operations Officer at IRCAI*

*Abstract*

Artificial Intelligence (AI) in Education has great potential for building more personalised curricula, as well as democratising education worldwide and creating a Renaissance of new ways of teaching and learning. We believe this is a crucial moment for setting the foundations of AI in education in the beginning of this Fourth Industrial Revolution. This report aims to synthesize how AI might change (and is already changing) how we learn, as well as what technological features are crucial for these AI systems in education, with the end goal of starting this pressing dialogue of how the future of AI in education should unfold, engaging policy makers, engineers, researchers and obviously, teachers and learners. This report also presents the advances within the X5GON project, a European H2020 project aimed at building and deploying a cross-modal, cross-lingual, cross-cultural, cross-domain and cross-site personalised learning platform for Open Educational Resources (OER).

*Authors*
**María Pérez-Ortiz,** University College London
**Erik Novak,** Jožef Stefan Institute
**Sahan Bulathwela,** University College London
**John Shawe-Taylor,** University College London

*November 2020*

## 1. Introduction

Education will be impacted, and possibly transformed, by AI. AI is already changing the knowledge and skills needed in our global and innovation centred world. But AI is also enabling innovative methods of teaching and learning [15]. This report offers a summary and some examples of how AI could support the task of learners and teachers around the world, reflecting on some of the social aspects of AI and the technical and pedagogical challenges of the field.

*AI in Education (AIEd) includes everything from AI-driven, personalized and conversational educative systems, intelligent agents, automatic scoring and assessment and learner-support chatbots, to AI-facilitated matching of learner to learner/teacher and collaborative learning, putting learners in full control of their learning [15].*

The field itself generates its own research questions: What is the nature of knowledge, and how can it be represented and understood with AI? How can a learner be supported in their lifelong learning path? How can a AI system support through personalisation via different styles of learning and be inclusive? How can the AI system be built in a transparent way so that it allows learners to self-reflect on their path and knowledge acquired? How can the AI system automatically gather and filter educational resources available online and suitable for the learner? Thus, it proves as well a research tool to test hypotheses within the learning sciences.

As a more specific example, AI can offer easy access to a complete individualization of learning in the form of your own lifelong learning companion. Particularly, formal evaluations have shown that intelligent tutoring systems, which use AI to support the tutoring process, produce similar learning gains as one-on-one human tutoring, which has the potential to increase student performance to around the 98 percentile in a standard classroom [25, 11, 2]. Additionally, intelligent tutors could effectively reduce by one-third to one-half the time required for learning [25], increase effectiveness by 30% as compared to traditional instruction [25, 14, 13], reduce the need for training support personnel by about 70% and operating costs by about 92%. These gains will enable great opportunities facilitating quality education in developing countries [18, 24].

## 2. Current challenges for AIEd

We identify several unique technical challenges within the application of AI to education. Many of these challenges are well-aligned with some of the progress in the X5GON project. While this list of challenges may not be complete, we believe that addressing these is vital in order to build a new generation of technological tools for education and that creating a dialogue around these would bring us closer to setting the foundations needed for the future of AIEd:

*Aim 1:*
*Scalable and automatic mining/comprehension of educational materials:* Automatic annotation and processing of educational materials has been a key challenge faced by different branches of the AIEd community since its inception. Many of the current applications rely simply on human labelling, but this is not scalable in practice. With groundbreaking ideas such as Open Education Resources emerging, the rate of production of novel educational resources has skyrocketed. However, learners can often be overwhelmed by all the available information, (specially as the number of choices increases), and could greatly benefit from an educational recommender. It is thus essential for these AIEd systems to understand the universal structure and direction of knowledge, identifying automatically knowledge prerequisites, topics covered and difficulty for educational materials, while at the same time filtering the materials by some metrics of quality assurance, all with the final goal of matching the most appropriate materials with the learners. This objective needs to be achieved across multiple languages and cultures, supporting automatic translations.

*Aim 2:*
*Understanding learners and personalisation:* AIEd systems need to be aware of the state of the learners, which involves keeping track of their ever evolving knowledge, interest and learning goals. Furthermore, these systems need to bring novelty and serendipity to the learner, to keep the user growing effectively and expose them to new high quality material they may not have found otherwise. These systems also need to consider learning preferences/styles and the special needs of users.

*Aim 3:*
**Transparency, conversation and Human-AI collaboration:** AIEd systems need to be transparent, to build trust and enable self-reflection in learners [8]. They need to keep the human in the loop to communicate the beliefs that the system holds about the user and allow the user to scrutinise the model parameters and the model's perception of the learner [1]. Such communication should be done in a conversational and intuitive way for the learner. This could be done for example by building chatbots that can interact with learners, helping them revise and explain concepts within the learner's knowledge gaps. Part of these aims not only apply to AI but more generally to the design of the user interfaces or the human-computer interaction. For example, by building visualisation of the materials in the user interface that allow the learner to understand/filter content more efficiently.

*Aim 4:*
**Extensive evaluation:** The hypotheses and techniques developed within the AIEd field need to be evaluated thoroughly and extensively, for example through the use of massive A/B testing, user studies, qualitative and quantitative learner feedback and available datasets. However, few educational datasets are publicly available, partly because the data comes from proprietary platforms that often do not release data that could be crucial to leveraging these personalised systems in the future. This is also in part due to privacy concerns, but this significantly precludes the advancement of the field. The community needs to propel the release of publicly available datasets, while finding ways to maintain the privacy of the users.

*Aim 5:*
**Generative and planning tasks:** AIEd systems need to conquer as well highly ambitious generative tasks, such as automatic generation of test questions and exercises, as well as personalised learning paths that entail a large planning component.

*Aim 6:*
**Data efficiency and scalability:** Such systems need not to overload the user and be able to infer information from learners simply from interactions with the educational materials, rather than relying heavily on explicit feedback. Finally, such systems need to have a moderate computational complexity, in order to allow for thousands of learners concurrently.

## 3. X5GON

Open Educational Resources (OERs), defined as teaching, learning and research material available in the public domain or published under an open license [23], are growing at a very fast pace. While OERs could be the richest collection of learning resources that will ever be available freely to all people, a challenge that remains to be addressed is to identify ways to enhance adoption, ease of use and discovery of these OERs. **The project <u>X5GON: Cross Modal, Cross Cultural, Cross Lingual, Cross Domain, and Cross Site Global OER Network</u> is an initiative that attempts to develop various innovative and open technology elements that will converge the currently scattered OERs.**

When considering the OER ecosystem, the rapid rate of creation of new resources [19], together with the imbalanced number of materials coming from different languages/cultures [17], the variable quality of resources [9], and the clustering of resources in multiple isolated silos [21] are some of the noteworthy challenges that should be surmounted. X5GON combines various elements such as content understanding, user modelling and personalisation, and quality assurance to create a unified network of OERs across the globe. The open technologies that are anticipated to be developed through X5GON envisages access to OERs focusing on the 5 Xs, in a 1) Cross Modal, 2) Cross Cultural, 3) Cross Lingual, 4) Cross Domain, and 5) Cross Cultural setting.

*X5GON project has pursued several fronts to address the previously mentioned challenges.* For example, gaining visibility into the majority of educational resources available in the world, as a prerequisite to unlocking their potential. X5GON is comprised of several core technologies:

*1. X5GON Connect Service:*
Connecting OER repositories across the world into a single network. Within the network, the Connect Service [27] is able to identify how a user is navigating between OERs on the same or across other repositories, thus enabling user pattern analysis. [20, 22]

*2. X5GON Discovery:*
AI-powered search engine [28] that enables learners to discover learning materials. It enables finding OERs by type, license, and language. In addition, the X5GON team partnered with Creative Commons and integrated their Creative Commons Catalog API [10], thus supporting searching through CC licensed images.

*3. X5GON Recommend:*
Aligning learning materials with learner context by finding other OERs they are likely to be interested in. The recommendation plugin is a component that allows users to find materials that are related to the OER they are currently observing.

*4. X5GON Translate:*
*Providing high quality automatically generated t*ranslations of learning materials beyond borders. This is achieved through the use of the Media Transcriptions and Translation Platform [12], supporting several languages and focusing the development of high quality translation models for minority languages.

*5. X5Learn:* An intelligent learning platform that bridges the right educational resource to the right user at the right time within a futuristic user interface [3].

*6. X5GON TrueLearn:* A scalable, transparent educational recommender for lifelong learners [5, 4].

*7. X5GON Moodle:* A Moodle plug-in to use X5GON technologies to build learning activities.

*8. X5GON Blind:* A prototype learning environment for the blind and visually-impaired [26]. The environment leverages existing technologies adapted for the visually-impaired and integrates X5GON TrueLearn to provide recommendations [16].

*9. X5GON Feed:* Providing quality data to further research in open education. The public X5GON API [29] enables accessing OER metadata, transcriptions and translations, as well as the models and services described above.

*10. X5GON datasets:* X5GON has released several datasets that can be used to approach some of the previously mentioned technical challenges [6, 7]. As the project approaches its end more datasets will be released.

4. Discussion
AI will impact the education industry and society greatly and broadly. However, there are enormous technical and pedagogical challenges ahead. This report has summarised some of these challenges and the foundations that the X5GON project has set towards these. While plenty remains to be done, we believe that building a dialogue around the future and challenges of AIEd will bring us closer to a next generation of personalised technological educative systems with great potential across the globe.